\newcommand{\beq}{\begin{equation}}
\newcommand{\eeq}{\end{equation}}

\newcommand{\Xmax}{X_{\rm max}}
\newcommand{\Xmumax}{X^{\mu}_{\rm max}}
\newcommand{\gcm}{{\rm g\,cm^{-2}}}

\documentclass[3p,times,twocolumn]{elsarticle}

\usepackage{ecrc}

\volume{00}

\firstpage{1}

\journalname{CRIS2015, submitted to Nuclear and Particle Physics Proceedings}

\runauth{L. Cazon for the Pierre Auger Collaboration}

\jid{nuphbp}

\jnltitlelogo{Nuclear Physics B Proceedings Supplement}

\usepackage{amssymb}

\usepackage[figuresright]{rotating}

\begin{document}

\begin{frontmatter}



\dochead{}

\title{Hadronic physics with the Pierre Auger Observatory}


\author{L. Cazon\fnref{a}}
\author{for the Pierre Auger Collaboration\fnref{b}\corref{*}} 

\address[a]{LIP: Laborat\' orio de Instrumenta\c c\~ ao e F\' \i sica Experimental de Part\'\i culas, Av. Elias Garcia 14-1, 1000-149 Lisbon, Portugal }
\address[b]{Observatorio Pierre Auger, Av. San Mart\' \i n Norte 304, 5613 Malarg\" ue, Argentina}
\cortext[*]{e-mail: auger\_ spokespersons@fnal.gov}\cortext[*]{Full author list: http://www.auger.org/archive/authors\_ 2015\_09.html}

\begin{abstract}
Extensive air showers are the result of billions of particle reactions initiated by single cosmic rays at ultra-high energy.  Their  characteristics are sensitive both to the mass of the primary cosmic ray and  to the fine details of hadronic interactions.  Ultra-high energy cosmic rays can be used to experimentally extend our knowledge on hadronic interactions in energy and kinematic regions beyond those tested by human-made accelerators.

We report on how the Pierre Auger Observatory is able to measure the proton-air cross section for particle production at a center-of-mass energy per nucleon of 39 TeV and 56 TeV and also to constrain the new hadronic interaction models tuned after the results of the Large Hadron Collider, by measuring:  the average shape of the electromagnetic longitudinal profile of  air showers, the moments of the distribution of the depth at which they reach their maximum, and the content and production depth of muons in air showers with a primary center-of-mass energy per nucleon around and above the 100 TeV scale.

\end{abstract}


\end{frontmatter}


\section{Introduction}
\label{Introduction}

Particle reactions beyond the energies reached by the Large Hadron Collider (LHC) occur every second at the top of the Earth's atmosphere. When an ultra-high energy cosmic ray (UHECR) collides with an air nucleus, it initiates an extensive air shower (EAS) of secondary particles that produce detectable electromagnetic radiation at different wavelengths due to interactions with the atmosphere. Eventually, a sizable fraction of those particles reach the ground surface, and can also be detected.

The Pierre Auger Observatory is capable of detecting 
 EAS by means of the Fluorescence Detectors (FD) which measure the fluorescence light emitted during the shower development in the atmosphere, and also the particles reaching the ground by means of the Surface Detector (SD), which is an extensive array of water-Cherenkov detectors (WCD) occupying an area of 3000 km$^2$  in the Argentinian high plateau.
A more general description can be found in \cite{ThePierreAuger:2015rma} and \cite{CRISCoutu} (in these proceedings) and references therein.

The mass of the cosmic ray primaries is a key ingredient needed to solve the long-standing mystery on the origin of UHECRs. Due to the hadronic nature of the primaries, shower observables share phase-space between the different possibilities for the primary mass and the uncertainties on the interaction models 
at the highest energies and forward region.

The aim of this contribution to CRIS2015 was to present the techniques and measurements made by the Pierre Auger Observatory which are relevant to constrain hadronic interactions. In \cite{Ostapchenko2,Werner,Anh} and \cite{Ostapchenko,Pierog} there is the set of references  used within this paper for the high energy hadronic models prior and after the LHC.

The paper is organized as follows: in section 2 the hadronic and electromagnetic (EM) components of air showers are briefly introduced,  and their internal dynamics are described. In section 3, measurements related to the EM cascade are described, namely: the moments of the distribution depths of shower maximum, the shape of the average shower development profile, and finally the proton air cross-section. In section 4, measurements related to the hadronic cascade through muons are addressed, namely: the muon content in inclined showers, the muon production depth, and measurements of the surface water cherenkov detector response. Finally, in section 5 the main conclusions are presented. 

\section{The internal air shower dynamics}

In an extensive air shower initiated by a proton at $10^{19}$ eV, after the first interaction $\sim$ 80\% of
the produced particles are pions and $\sim$ 8\% are kaons. Neutrons/protons
are produced with an overall probability of $\sim$ 5\%,
and the rest is shared among other particles at the
sub-percent level as given by QGSJETII-03 on average \cite{Cazon:2013dm}.

One third of pions are neutral, and decay almost immediately, feeding the EM cascade, whereas the rest are charged pions, which either interact (sustaining
the multiplicative process of the cascade), or decay into muons.
At the highest energies, all kaons keep interacting  hadronically. When the critical energy of the different kaons is reached, they feed the EM and hadronic component through neutral or charged pions or directly decay into muons, according to their specific branching ratios \cite{Cazon:2013dm}. 
 Finally, neutrons and protons keep interacting hadronically.

Most of the energy carried by the secondaries after the first interaction undergo new hadronic interactions, sustaining what is thus called the {\it hadronic cascade} until it eventually vanishes, leaving muons as measurable trace.

On the other hand, a fraction  of the energy ($\sim$ 30\%) is subtracted in each reaction from the hadronic cascade into the electromagnetic cascade
 After typically 2 hadronic generations, most of the energy has been transferred to the EM cascade, and since there is very little feedback to the hadronic cascade, mostly by photopion production, both hadronic and EM cascades can be considered decoupled.

 Details about multiparticle production are linked to the shower observables.  Among the most important ones is the ratio of hadronic to EM particle production, which affect the hadronic (EM) cascade through substracting (injecting) energy at the different stages of development.

\section{The EM cascade}
\subsection{The moments of shower maximum}

\begin{figure}[h!]
  \begin{center}
    \includegraphics[width=7cm]{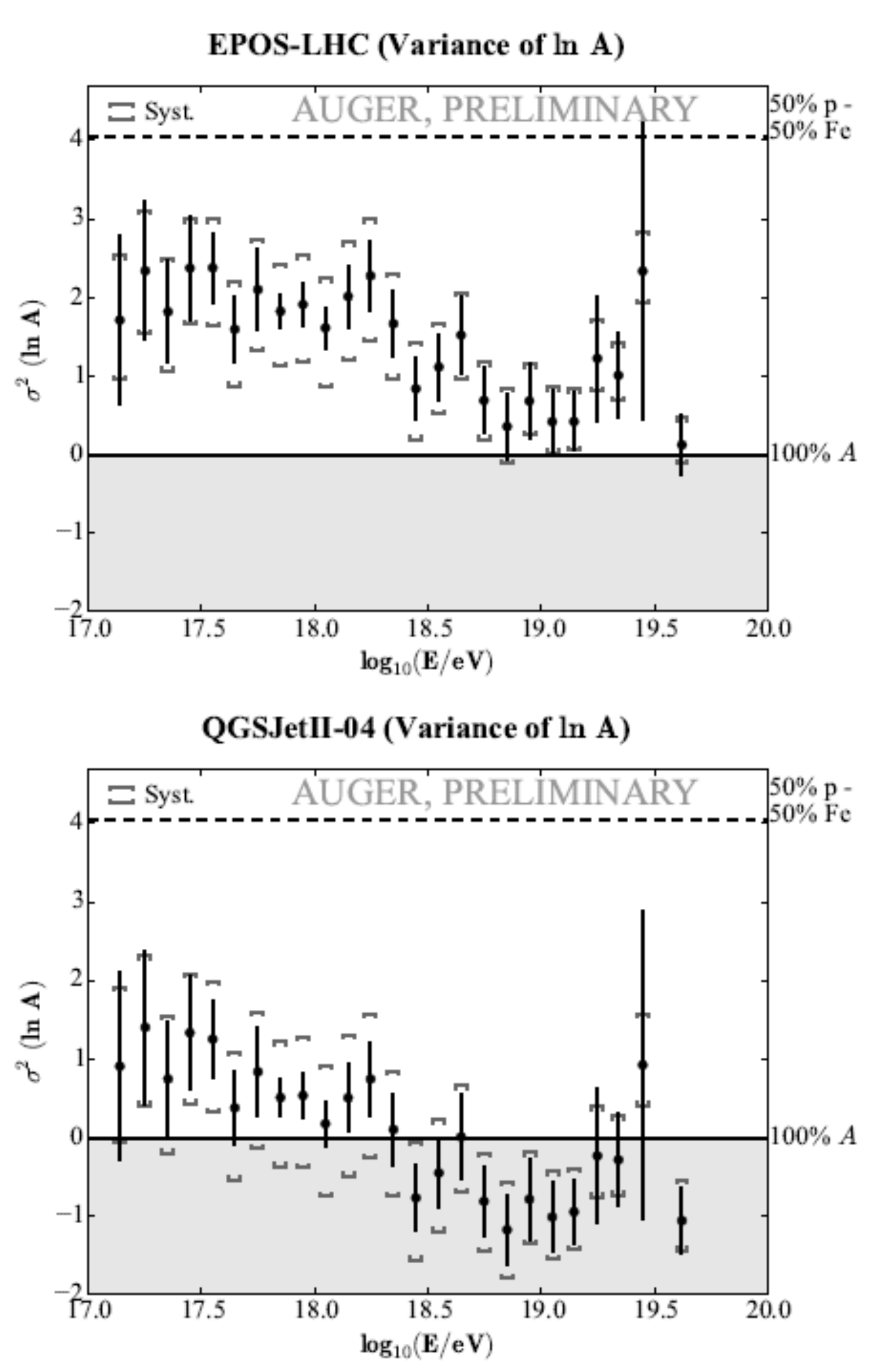}
\caption[]{Variance of $\ln A$ according to two different hadronic models.}
\label{f:sigmalnAmoments}
 \end{center}
\end{figure}

The depth at which the number of  charged particles reaches its maximum in the air shower, $\Xmax$, is sensitive to the mass composition of the primaries through the depth of the first interaction, related to the nuclei-air cross-section and also to the posterior shower development. These charged particles are completely dominated by electrons belonging to the EM cascade. Experimentally, the longitudinal profile can be measured using fluorescence light emitted by atmospheric nitrogen molecules excited by the passage of charged particles. At the Pierre Auger Observatory, which is continuously taking data since January 2004, such measurements are performed using the FD consisting of 24 telescopes placed at 4 locations and since June 2010 using the High Elevation Auger Telescopes (HEAT).

 The first two moments of the $\Xmax$-distribution $\left< \Xmax \right>$ and $\sigma[\Xmax]$ are related to the first two moments of the distribution of the logarithm of masses of primary particles $\left< \ln A \right>$ and $\sigma^2[\ln A]$ as 
\begin{eqnarray}
\langle \Xmax \rangle &=& \langle \Xmax \rangle_p + f_E \langle \ln A \rangle \\
\sigma^2[ \Xmax] &=& \langle \sigma^2_{sh} \rangle + f^2_E \sigma^2[\ln A]
\label{eq:sigmalnA}
\end{eqnarray}

where $\langle \Xmax \rangle_p$ is the mean $\Xmax$ for protons and $\langle \sigma^2_{sh} \rangle = \Sigma f_i \sigma^2_i[\Xmax]$ is the composition-averaged shower-to-shower fluctuations. $f_E$ depends  on the details of hadronic interactions, and in practice is parametrized from the interaction models \cite{Abreu:2013env}. See \cite{Aab:2014kda} for a complete discussion and the latest update of results in \cite{ICRCPorcelli}.
 Figure \ref{f:sigmalnAmoments} displays the results for $\sigma^2[\ln A]$ for the two post-LHC models. ($\langle \ln A \rangle$ can be seen in Fig. \ref{lnAupdated}, which is relevant for discussion in section 4.2). For QGSJETII-04 the value of $\sigma^2[\ln A]$ above $E > 10^{18.3}$ eV reaches negative values beyond systematic uncertainties. This imposes constraints on the description of the $\Xmax$ moments by QGSJETII-04: the composition-averaged shower-to-shower fluctuations predicted by the model exceed the measured $\Xmax$ fluctuations.

\subsection{Measurement of the p-air cross section}

\begin{figure}[h!]
  \begin{center}
    \includegraphics[width=7cm]{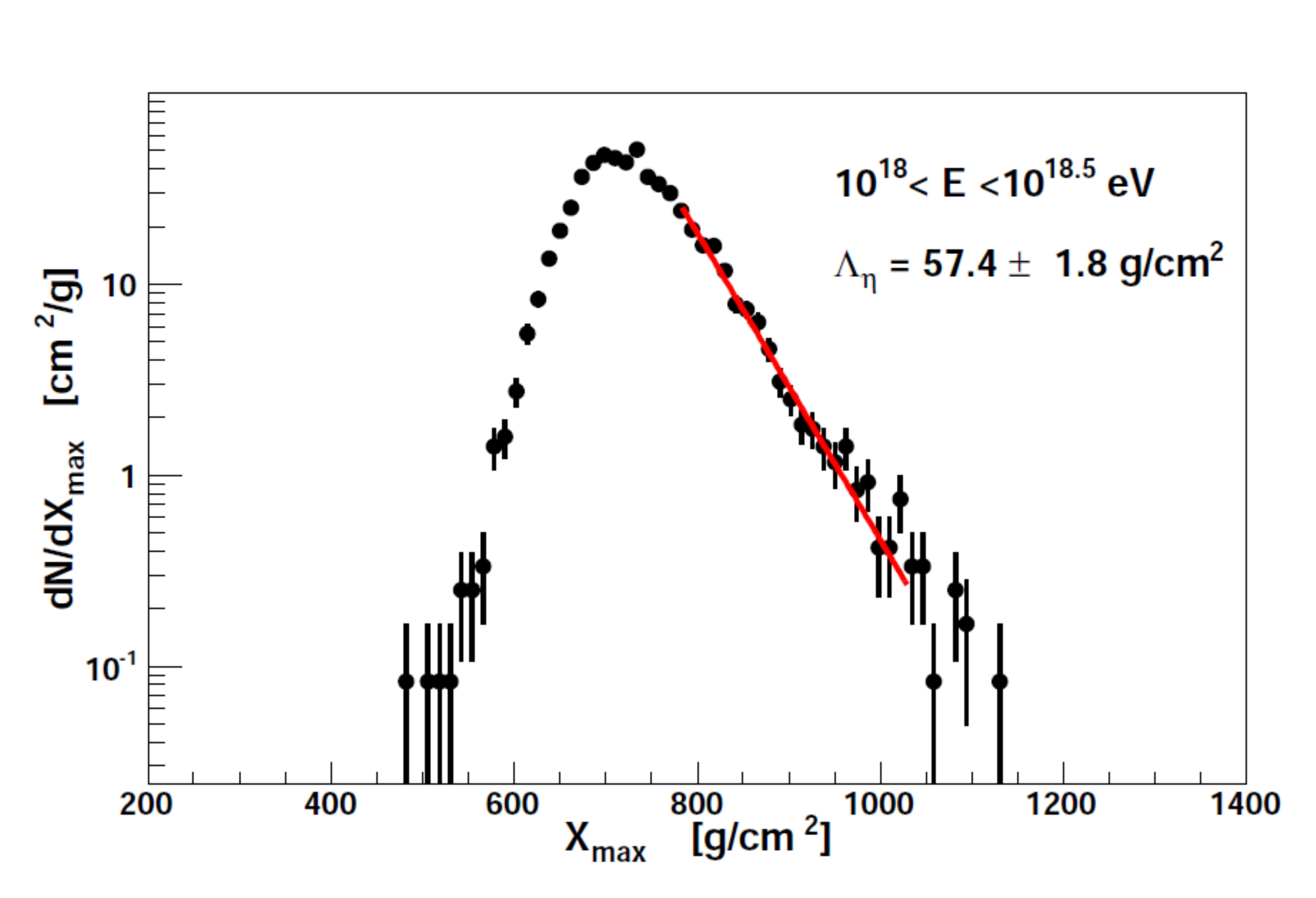}
\caption[]{$\Xmax$-distribution in the highest energy bin and the result of the fit to the exponential tail. See \cite{Collaboration:2012wt} and \cite{ICRCRUlrich}  for details.}
\label{XmaxHighE}
 \end{center}
\end{figure}

\begin{figure}[h!]
  \begin{center}
    \includegraphics[width=7cm]{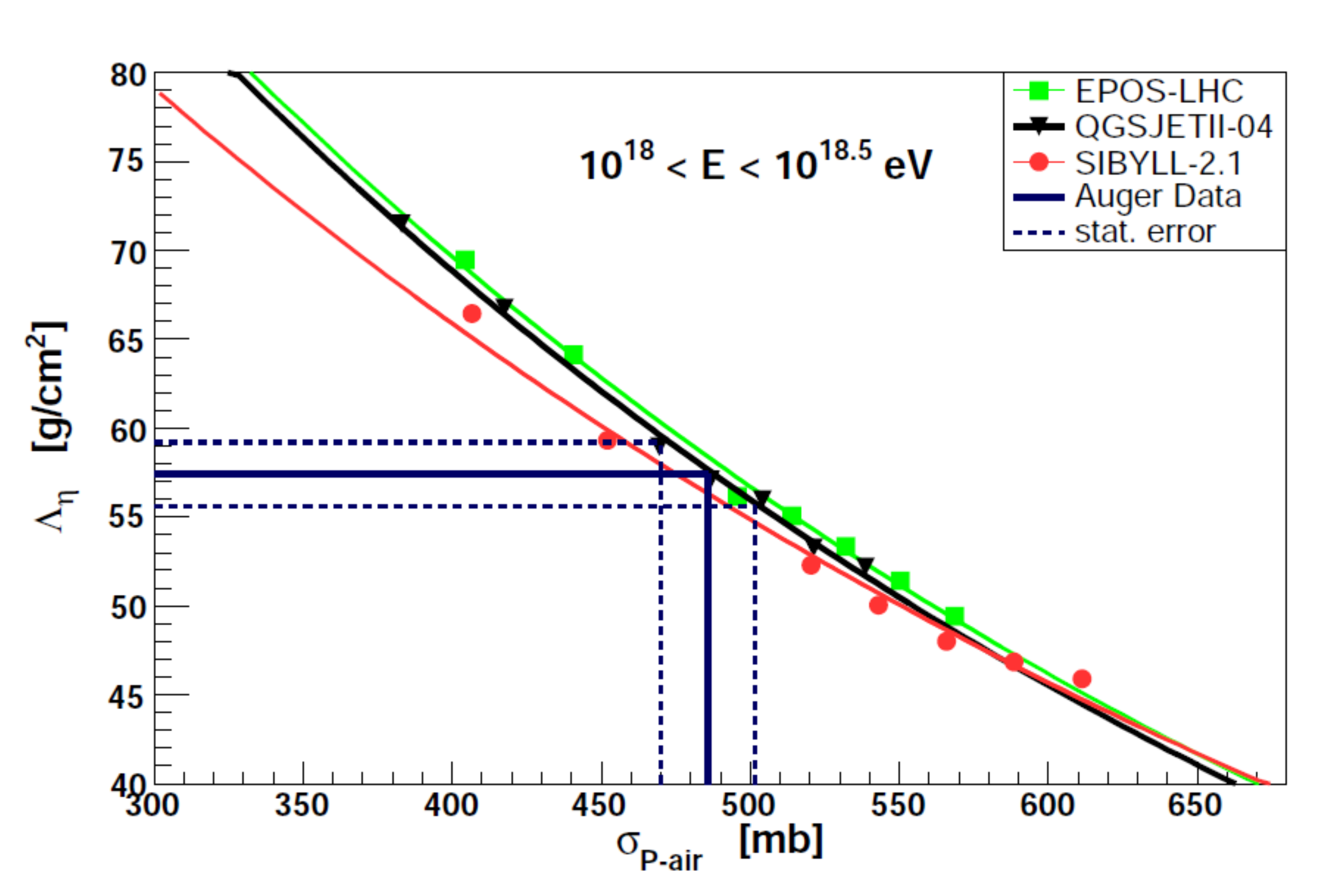}
\caption[]{Conversion from $\Lambda_{\eta}$ to $\sigma_{\rm p-air}$ at the highest energy bin. The simulation included all detector resolution effects, while the data is corrected for acceptance effects. The solid and dashed lines show the $\Lambda_{\eta}$ measurement and its projection to $\sigma_{\rm p-air}$ using the average of all models. See \cite{Collaboration:2012wt} and \cite{ICRCRUlrich} for details.}
\label{lambdahigh}
 \end{center}
\end{figure}

\begin{figure}[h!]
  \begin{center}
    \includegraphics[width=7cm]{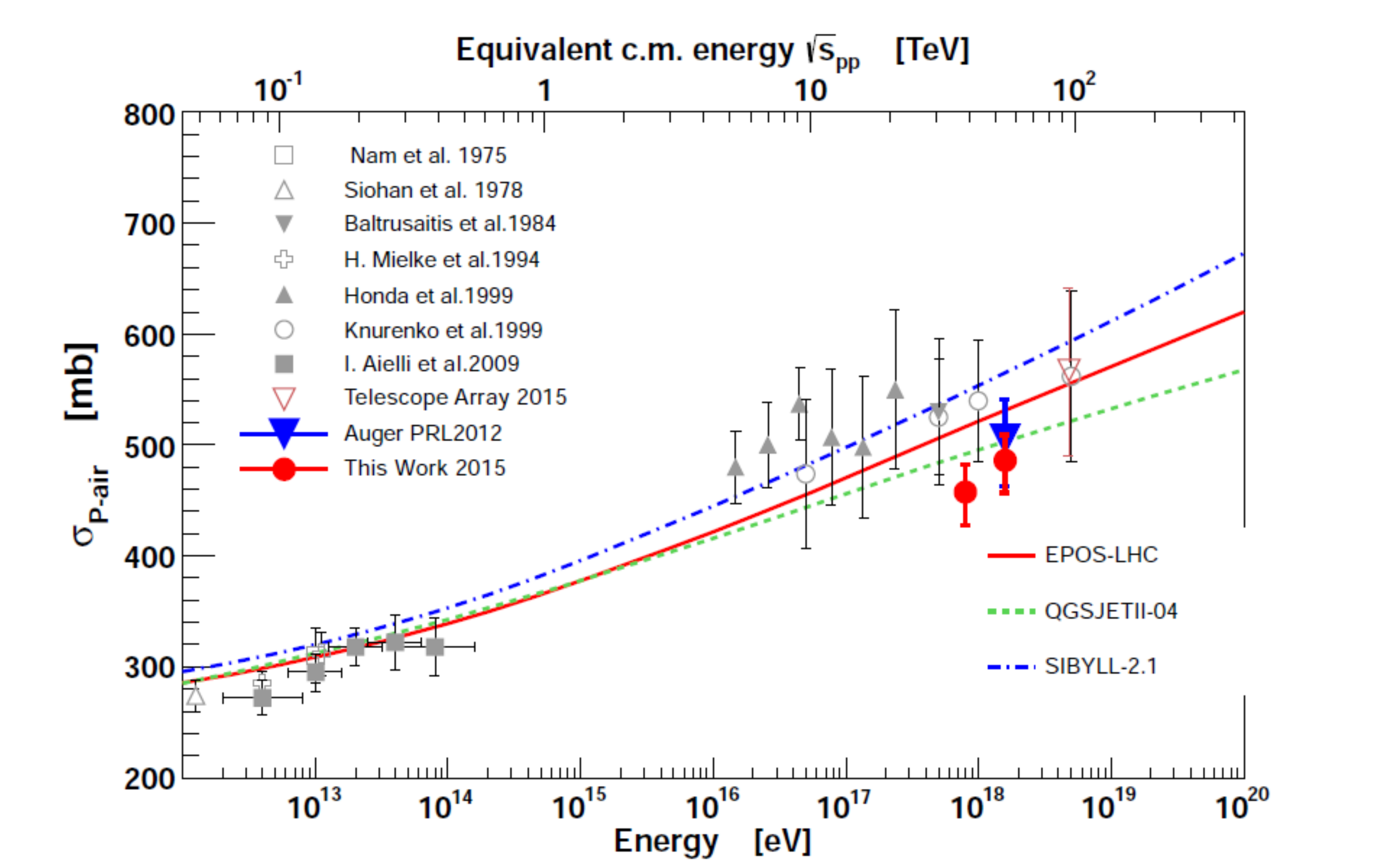}
\caption[]{The $\sigma_{\rm p-air}$ measurement compared to previous data and model predictions. See \cite{Collaboration:2012wt} and \cite{ICRCRUlrich} for references.}
\label{cx}
 \end{center}
\end{figure}

The $\Xmax$-distributions also contain information about the cross section of the primary with the air. Among all primaries, proton has the smallest cross-section and therefore the maximum interaction length.  The primary interaction is reflected into the $\Xmax$-distribution by an $\exp(-\Xmax/\Lambda_\eta)$ dependency at highest $\Xmax$ values, being $\Lambda_\eta$ a constant directly related to the primary interaction length. By selecting the deepest showers and ensuring  enough proton presence, it is possible to measure the proton-air cross section. Details about the method can be found at \cite{Collaboration:2012wt}. In \cite{ICRCRUlrich} this measurement has been updated with the post-LHC hadronic models and presented in two energy intervals: $\log_{10}(E/{\rm eV}) \in[17.8,18]$ and  $\log_{10}(E/{\rm eV})\in[18,18.5]$.
These energies were chosen so that they maximize the available events statistics and at the same time lie in the region most compatible with a significant proton fraction.  The averaged energy of the events in each bin transformed into primary center-of-mass energy per nucleon is
\begin{eqnarray}
_{{\rm Low\,}E{\rm \,bin\,\,}} \sqrt{s_{pp}}=38.7 {\rm \,\, TeV} \\
_{{\rm High\,}E{\rm \,bin\,\,}} \sqrt{s_{pp}}=55.5 {\rm \,\, TeV} 
\end{eqnarray}

In Fig. \ref{XmaxHighE} it can be seen the $\Xmax$-distribution and the results of the $\Lambda_\eta$ fit to the tail, for the highest energy bin.
  Fig. \ref{lambdahigh} shows the conversion into $\sigma_{\rm p-air}$, which gives the values
\begin{eqnarray}
_{{\rm Low\,}E{\rm \,bin\,\,}}\sigma_{\rm p-air}&=&458\pm18(stat)^{+19}_{-25}(sys) {\rm \,\, mb} \\
_{{\rm High\,}E{\rm \,bin\,\,}}\sigma_{\rm p-air}&=&486\pm16(stat)^{+19}_{-25}(sys) {\rm \,\, mb} 
\end{eqnarray}

While the composition of primary cosmic rays in the energy range under investigation is compatible with being dominated by protons \cite{Aab:2014aea}, a contamination with Helium cannot be excluded up to 25\%, and it is one of the main sources of systematics.  Particles heavier than Helium have only negligible impact on the analysis. Photon contamination is excluded up to 0.5\%, and has also been accounted for in the systematic uncertainties. For a complete discussion on the systematic uncertainties refer to \cite{ICRCRUlrich}.

Results are plotted in Fig. \ref{cx} along with model predictions and different measurements, with good general agreement. The data might be  consistent with a rising cross section with energy, however, the statistical precision is not yet sufficient to make any claim.

\begin{figure}[h!]
  \begin{center}
    \includegraphics[width=7cm]{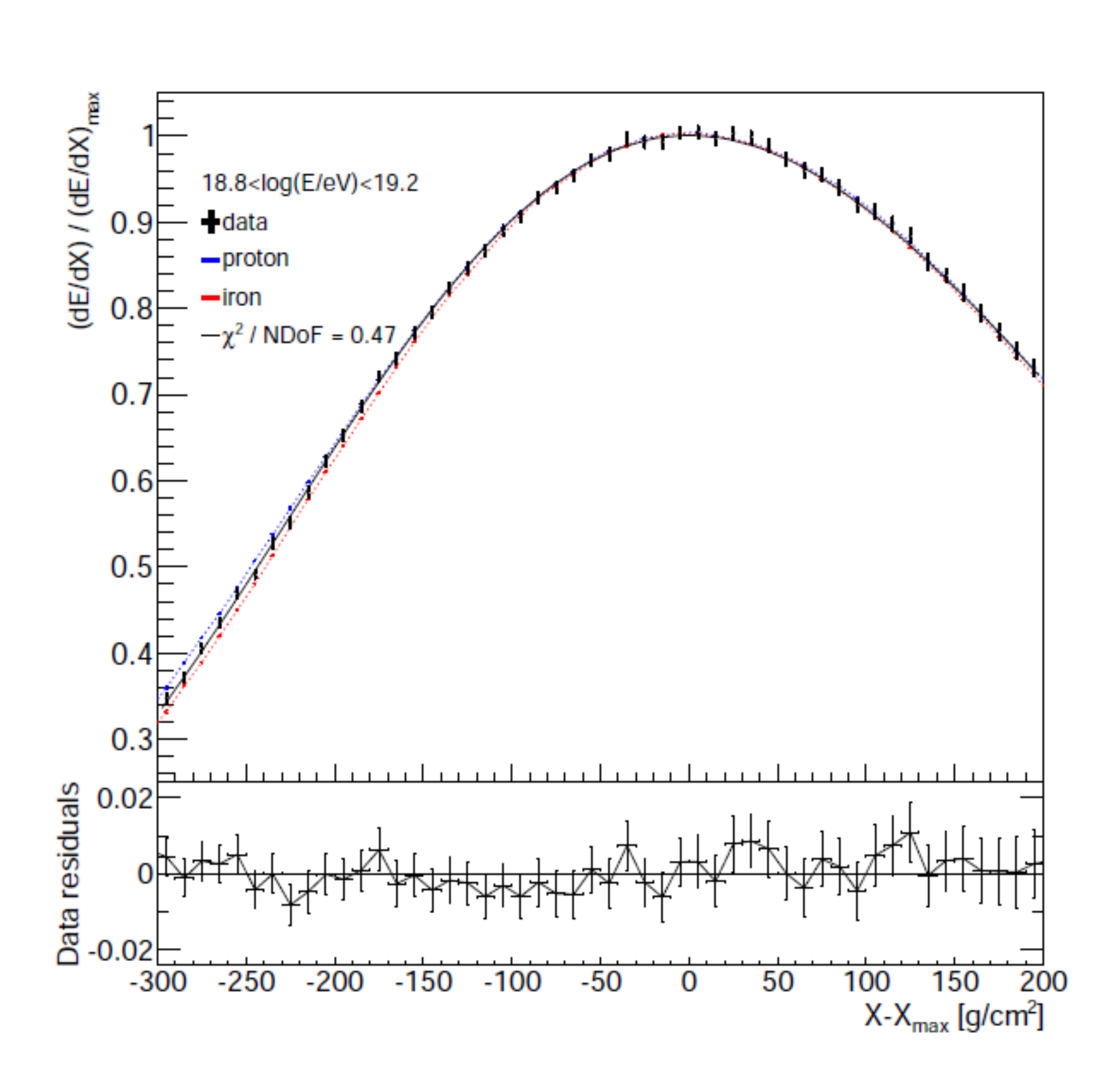}
\caption[]{Average profiles for energies between $10^{18.8}$ eV  and $10^{19.2}$ eV within the fitting range $X-\Xmax \in [-300,200]\,\gcm$. Data is shown in black, proton and iron reconstruction in blue and red respectively, with the Gaisser-Hillas fit superimposed. 
The residuals on the fit to the data are shown in the bottom. }
\label{profiles}
 \end{center}
\end{figure}

\subsection{Average shower profile}
A Gaisser-Hillas function (GH), typically used to describe the EM longitudinal profiles can be rearranged and written in the form
\beq
\frac{dE}{dX}=\left(1 + R \frac{X-\Xmax}{L} \right)^{R^{-2}} \exp\left(-\frac{X-\Xmax}{RL} \right)
\label{eq:USP}
\eeq
where  $L$ and $R$ are related to the standard GH parameters as $L=\sqrt{|X_0-\Xmax|\lambda}$ and $R=\sqrt{\lambda/|X_0-\Xmax|}$.
These parameters have the new meaning of a width ($L$) and an asymmetry ($R$) with respect to a Gaussian distribution, and  have been shown to be sensitive both to the high energy hadronic models and to the mass composition of the primary \cite{Andringa}\cite{Lipari} in an event by event basis, making the most of the parameter correlation ($\Xmax, R, L$), and in averaged showers \cite{Conceicao:2015toa}. 

High quality FD events have been centered around $\Xmax$ to obtain the average shower profile, and fitted to equation \ref{eq:USP} within the range $X-\Xmax \in [-300,200]\,\gcm$. 
Figure \ref{profiles} displays the averaged profiles  for data and simulations around $10^{19}$ eV. Details of the whole procedure can be found in \cite{ICRCFDiogo}. 

\begin{figure*}[th!]
  \begin{center}
    \includegraphics[width=13cm]{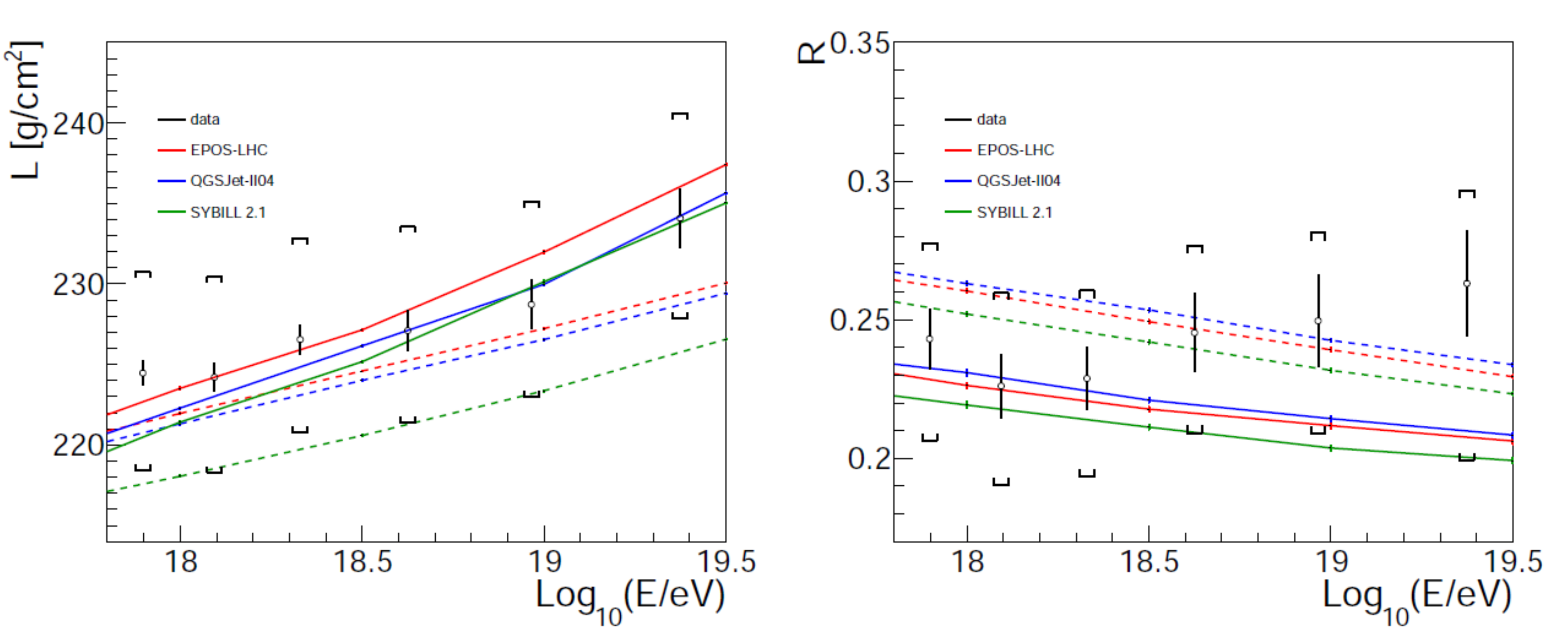}
\caption[]{$L$ (left) and $R$ (right) as a function of energy. The data is shown in black, with the vertical line representing the statistical error and the brackets the systematic uncertainty. Hadronic interactions models shown each with its color (see legend), with full lines being proton predictions and dashed lines the iron ones.}
\label{RLvsE}
 \end{center}
\end{figure*}

Figure \ref{RLvsE} displays $L$ and $R$ as a function of the energy for data and models. Both agree well with models within the current uncertainties. The  energy evolution  of the width $L$ is consistent with a linear increase with $\log_{10}(E/(\rm eV)$. The asymmetry, $R$, data displays an increase with energy not predicted by models, although it is contained within the systematic uncertainty of the measurement in absolute values.

\section{The hadronic cascade}
\subsection{Measurement of the muon production by means of inclined showers}

The EM component in inclined showers is largely absorbed in the atmosphere before reaching ground. Due to the effect of geomagnetic fields, the muon density observed at ground $\rho_{\mu}(\theta,\phi; {\bf r})$ is highly asymmetrical. It depends on the zenith angle ($\theta$) and azimuth angle ($\phi$). It has been shown that it is practically independent of the energy of the primary and  of the hadronic model, and can be described by
\beq
\rho_{\mu}(\theta,\phi; {\bf r}) \simeq N_{19} \rho_{\mu,19}(\theta,\phi;{\bf r}-{\bf r_c})
\label{eq:rho}
\eeq
where $\rho_{\mu,19}(\theta,\phi;{\bf r}-{\bf r_c})$ is an universal reference function for the number densities of muons expressed relative to the position of the core ${\bf r_c}$, and $N_{19}$ is a scale factor.
We  define $N_{\mu}$ as the surface integral of the l.h.s. of equation \ref{eq:rho}, the actual number of muons reaching ground, which can be calculated in Monte Carlo simulations. Equivalently, we define $N_{\mu,19}$ as the surface integral of the reference map $\rho_{\mu,19}(\theta,\phi;{\bf r}-{\bf r_c})$, and equation \ref{eq:rho} simply becomes $N_{\mu}\simeq N_{19} N_{\mu,19}$.  One can define the ratio $
R_{\mu}\equiv N_{\mu}/N_{\mu,19}$ and then, by fitting equation \ref{eq:rho} to real showers leaving $N_{19}$ as free parameters, one can use $N_{19}$ as estimation of $R_{\mu}$. This has been extensively tested with simulations, being the average difference between $N_{19}$ and $R_{\mu}$ for proton and iron simulated using QGSJET01, QGSJETII-04, and EPOS-LHC always below 5\%. To get an unbiased estimator, $N_{19}$ was corrected for the average bias of all the simulations to obtain an unbiased $R_{\mu}$. More details about this analysis can be found in \cite{Aab:2014pza}.
\begin{figure}[h]
  \begin{center}
    \includegraphics[width=7cm]{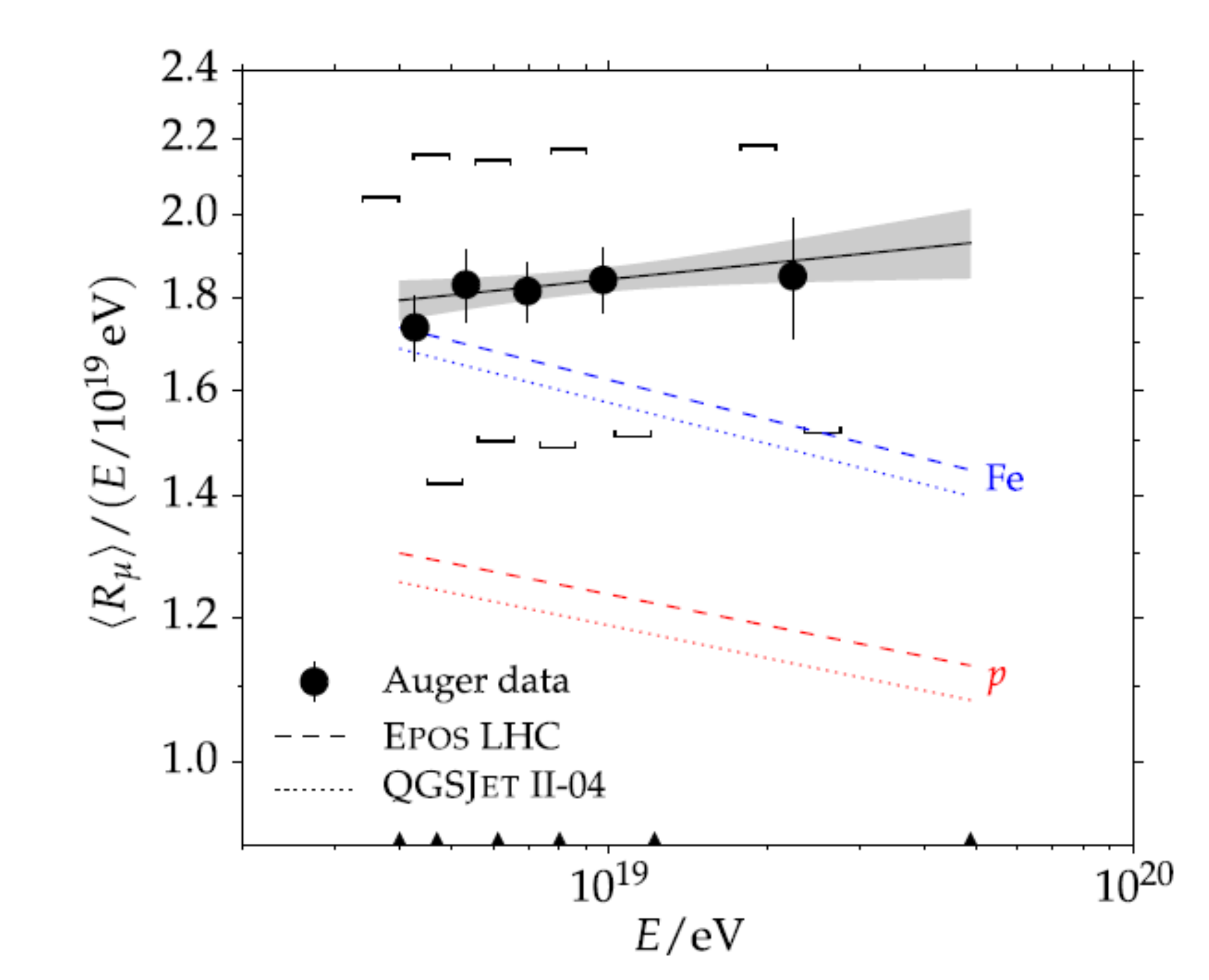}
\caption[]{$\langle R_\mu \rangle$ vs primary energy, compared to air shower simulations.}
\label{RmuE}
 \end{center}
\end{figure}

The average values of $R_{\mu}$ divided by the energy, are plotted for five energy bins in Fig. \ref{RmuE} and compared to simulations.
\begin{figure}[h]
  \begin{center}
    \includegraphics[width=7cm]{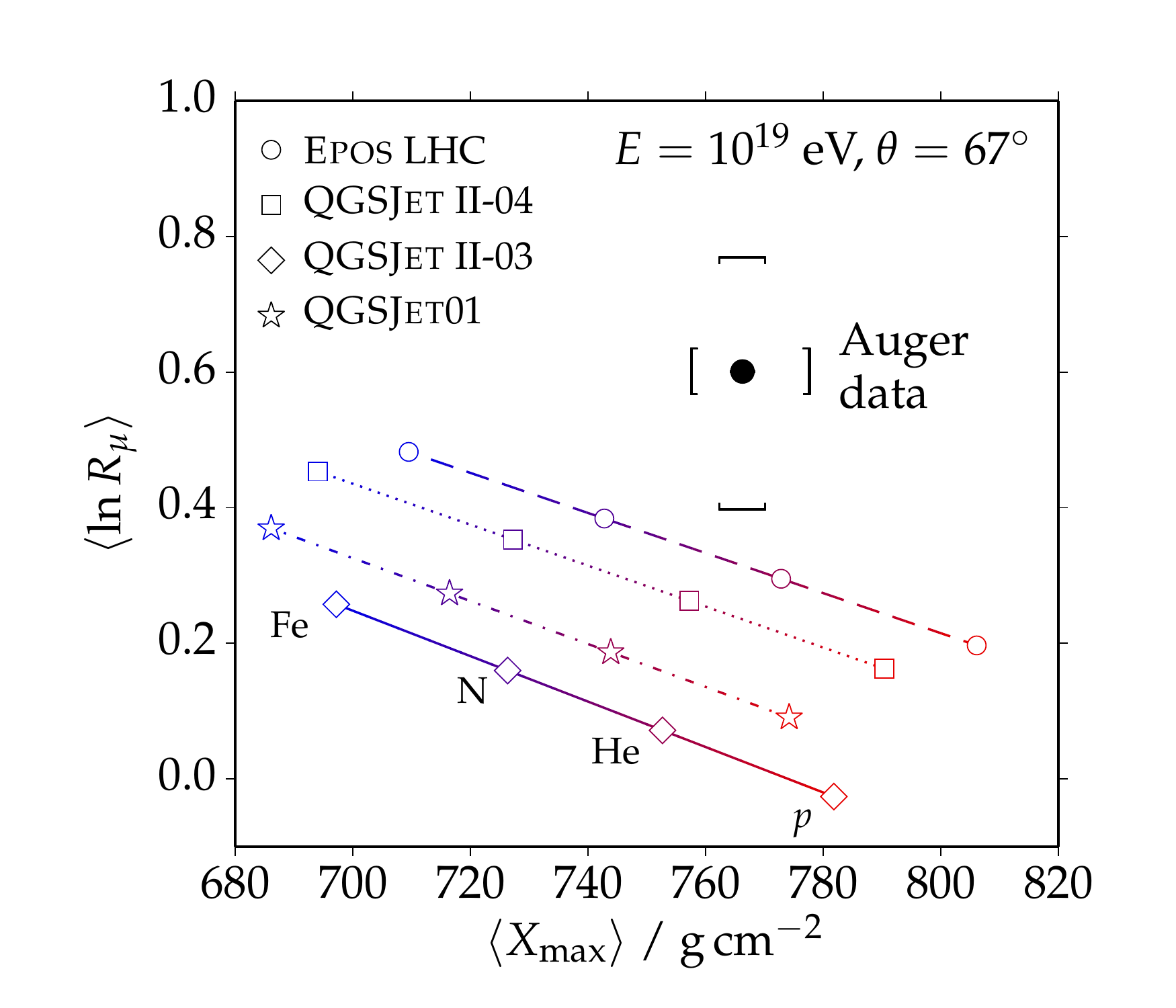}
\caption[]{$\langle R_\mu \rangle$ vs $\Xmax$ compared to air shower simulations.}
\label{RmuXmax}
 \end{center}
\end{figure}
Fig. \ref{RmuXmax} displays $\langle \ln R_\mu \rangle$ vs $\langle \Xmax \rangle$, with Auger  data and simulation data for different hadronic models and primaries.

\begin{figure*}[th!]
  \begin{center}
    \includegraphics[width=15cm]{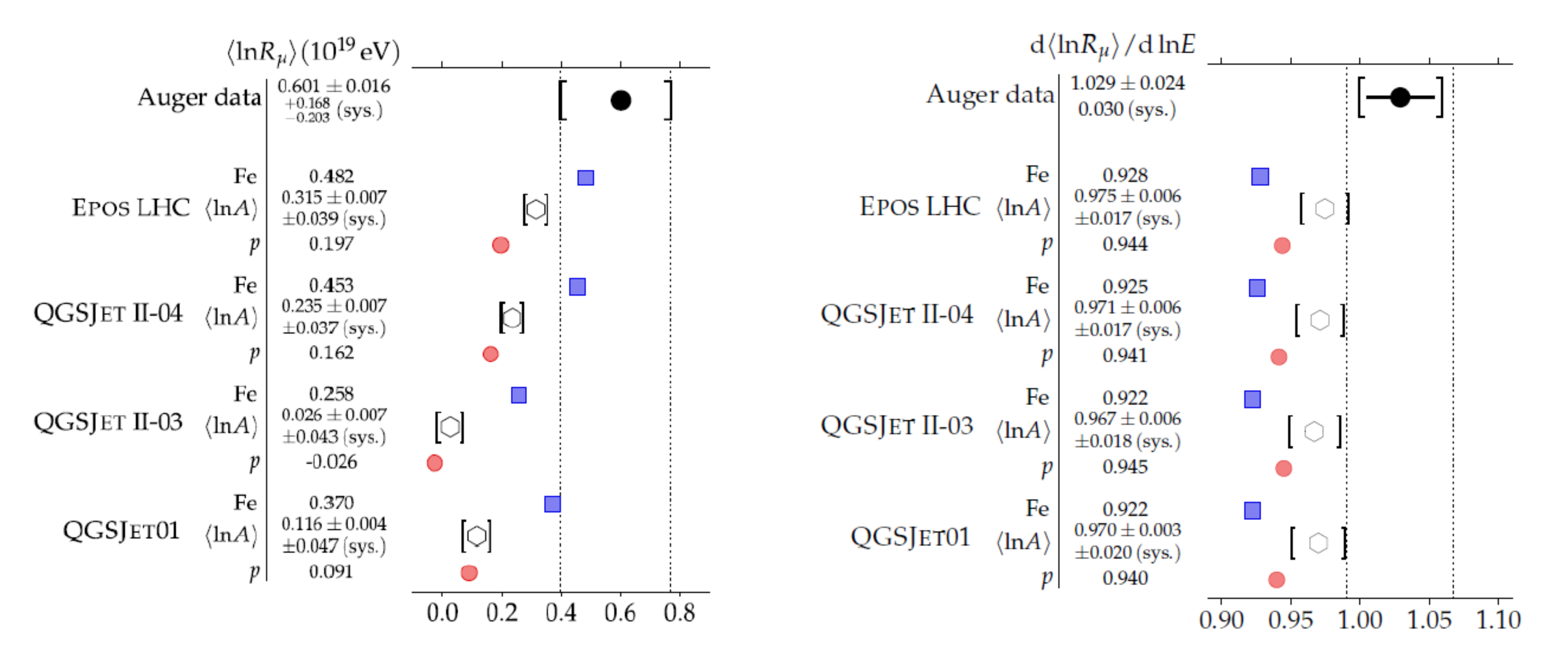}
\caption[]{Comparison of $\langle R_\mu \rangle$ at $10^{19}$ eV (left) and $d \langle R_\mu \rangle / d \ln E$ (right) with predictions for air shower simulations with different high energy hadronic models for a pure proton, pure iron and a mixed composition compatible with the FD measurements, labeled as $\langle \ln A \rangle$.}
\label{lnHAS}
 \end{center}
\end{figure*}

When we consider the values of $\langle \ln A \rangle$ deduced from $\langle \Xmax \rangle$, the measured values of $R_{\mu}$ indicate that the mean number of muons in the simulations have a deficit from 30 \% to 80 \% $^{+17}_{-20}$\% (sys)  at $10^{19}$ eV. Figure \ref{lnHAS} (left) summarizes the situation.
The logarithmic gain $d \langle \ln R_{\mu } \rangle / d \ln E$ (Fig. \ref{lnHAS} (right)) is  in agreement between all the models. Deviations of data from a constant proton (iron) composition  are observed at the level of 2.2 (2.6) $\sigma$. 

In \cite{ICRCFarrar}, an independent analysis confirming these results was performed with vertical showers, where both the muon and the EM components reach the SD. 
A possible overall energy scale shift in the FD reconstruction that could explain the muon discrepancy  was also excluded, confirming that models produce  a deficit in the hadronic cascade within the simulated showers.

\subsection{Measurement of the muon production depth}

In \cite{Cazon:2003ar} and \cite{Cazon:2012ti} it was shown that muons are produced within a narrow cylinder of a few tens of meters around the shower axis. After production, mainly coming from pion decay,  muons follow almost straight lines until they reach ground. Thus, their time delay with respect to a plane front traveling at the speed of light mainly depends on the length of the trajectories (geometric delay) with a second order correction stemming from subluminal velocities, which can be modeled from the average muon energy spectrum. It is thus possible to write a one-to-one map \cite{Cazon:2004zx} between the time delay with respect to a shower front $t$ and the production distance along the shower axis $z$ as:
\beq
z\simeq 
\frac{1}{2}
 \left( 
   \frac{r^2}
   {
     c (t-\langle t_{\epsilon} \rangle ) 
   }
   - c (t-\langle t_{\epsilon} \rangle )
 \right)
 + \Delta -\lambda_{\pi} 
\eeq
where  $\langle t_{\epsilon}\rangle$ is the {\it kinematic delay} due to subluminal velocities and $\lambda_{\pi}$ is a correction due to the pion decay length. $\Delta$ is  just the $z$-coordinate of the ground observation point in cylindrical coordinates, where the $z$-axis is on the shower axis, and the origin is on the core position.

The muon production depth (MPD) $X^\mu$  is obtained by integrating  the atmospheric density over the range or production distances. The MPD-distribution is derived adding all MPDs recorded in each of the SD stations of the event. A fit to the MPD-distribution with a Gaisser-Hillas function allows us to derive $\Xmumax$, the point in the shower development where production of muons reaches a maximum rate. 

In showers with a zenith angle between 55 and 65 degrees, a cut $r>1700$ m was introduced to further minimize the EM contamination and also the distortions introduced by the tank response into the time to depth mapping. Systematic uncertainty is  17 $\gcm$, and event by event resolution ranges from 90 $\gcm$ to 50 $\gcm$ from lowest energy to the highest energy range of applicability. A detailed description of the analysis can be found at \cite{Aab:2014dua}.

\begin{figure}[h]
  \begin{center}
    \includegraphics[width=7cm]{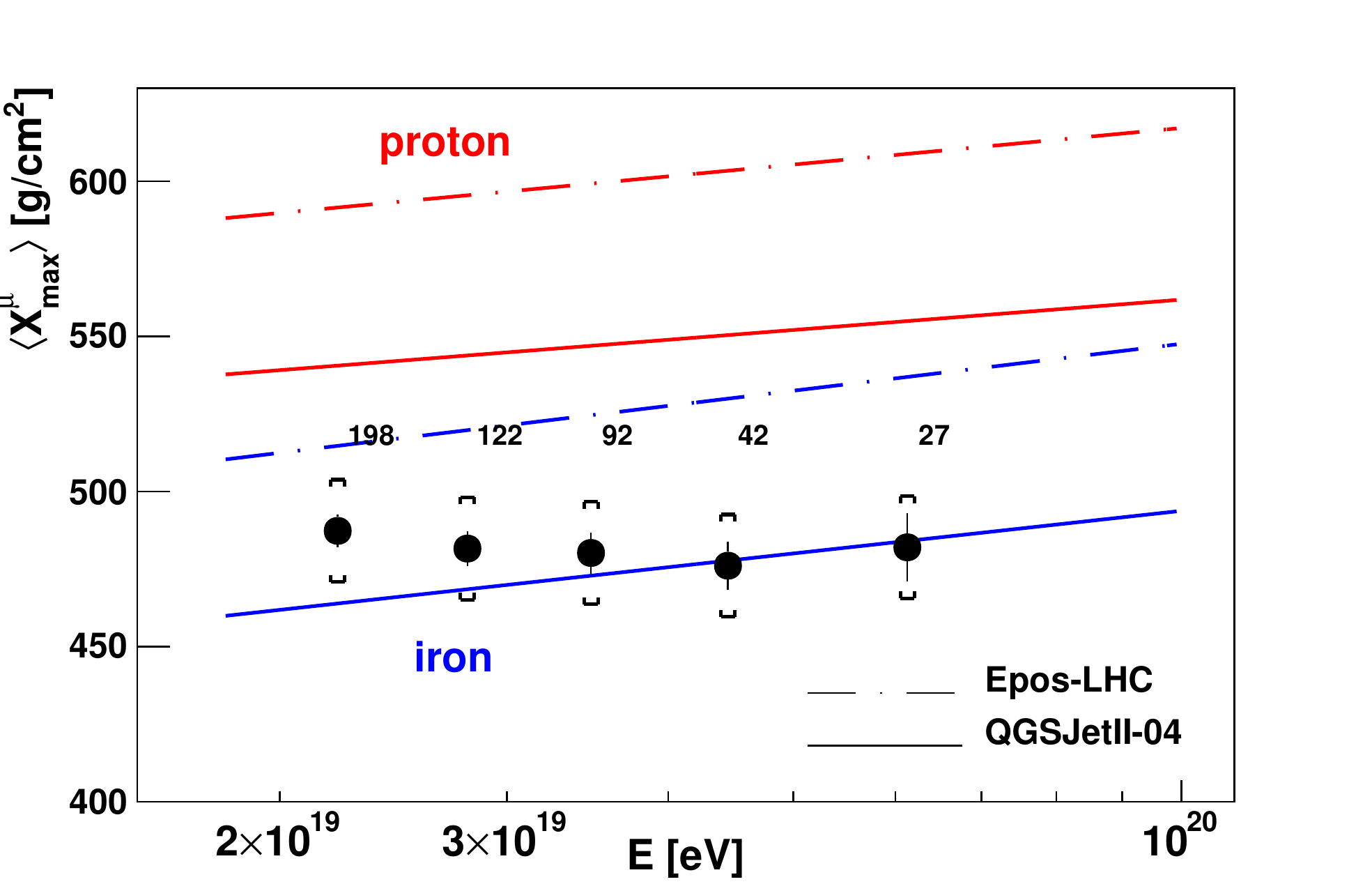}
\caption[]{Evolution of $\langle \Xmumax \rangle$ with energy. The number of events is indicated in each energy bin, and brackets indicate the systematic uncertainties.}
\label{ERmpd}
 \end{center}
\end{figure}

The evolution of the measured $\langle \Xmumax \rangle$ with energy is shown in Fig. \ref{ERmpd} along with the two post-LHC models. Both predict a similar evolution of $\langle \Xmumax \rangle$ for proton and iron but a considerable difference in its absolute value. While Auger data are bracketed by QGSJETII-04, they fall below EPOS-LHC prediction for iron, thus demonstrating the power of the MPD analysis to constrain hadronic interaction models.

\begin{figure*}[th!]
  \begin{center}
    \includegraphics[width=12cm]{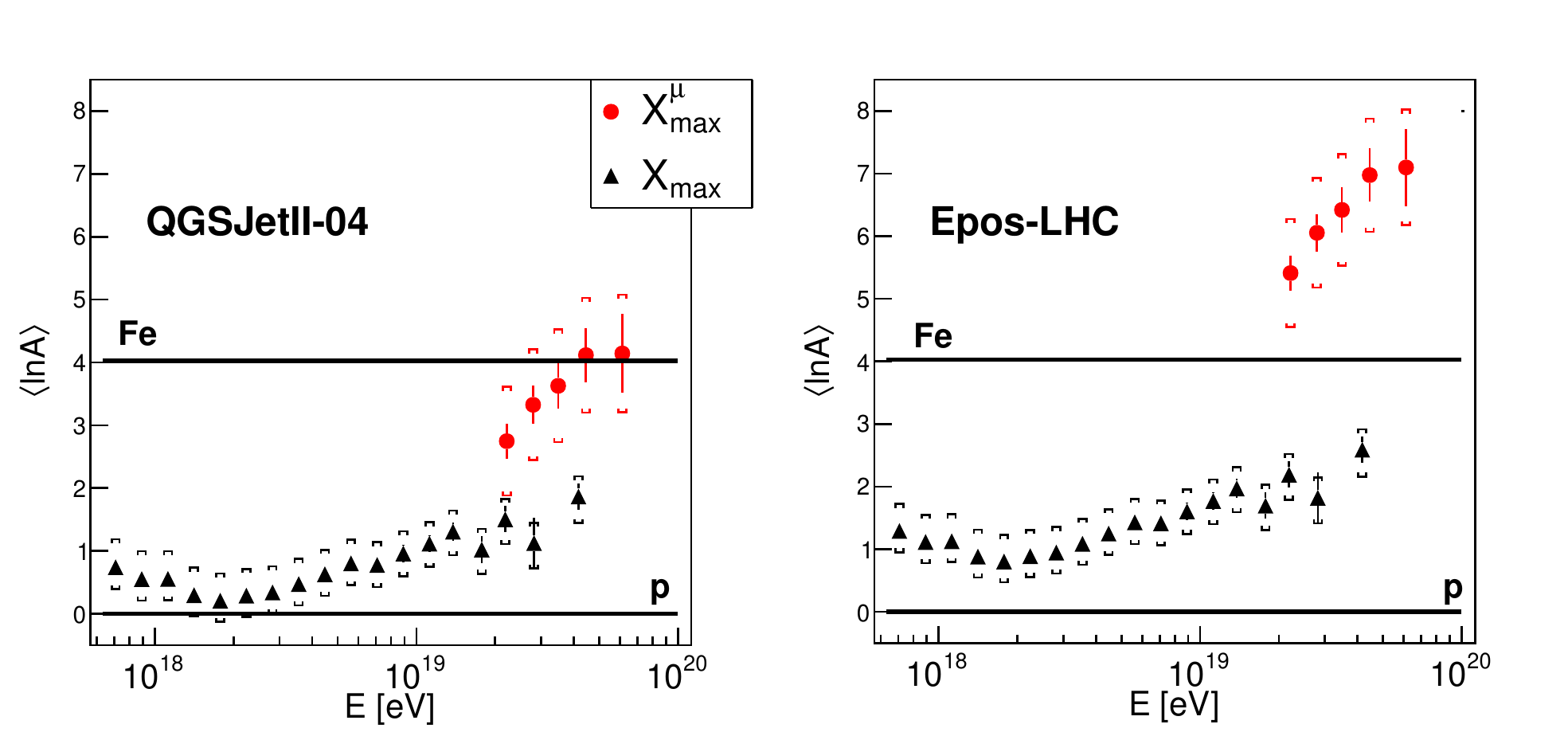}
\caption[]{$\langle \ln A \rangle$ from $\Xmax$(triangles) and $\Xmumax$ (circles) as a function of the energy for both post-LHC models.}
\label{lnAupdated}
 \end{center}
\end{figure*}

Both $\langle \Xmax\rangle$ and $\langle \Xmumax \rangle$ can be linearly converted into $\langle \ln A \rangle$ for each one of the models. Figure \ref{lnAupdated} shows the up-to-date results of this comparison \cite{ICRCCollica}. For QGSJETII-04, $\langle \ln A \rangle$ values are within 1.5 $\sigma$, while for the case of EPOS-LHC the results from $\langle \Xmumax \rangle$ indicate primaries heavier than iron and the EM and hadronic predictions are incompatible at a level of at least 6 $\sigma$.

At closer distances to the core and in more vertical showers, the signal in the WCD  is made of different muonic and EM contributions. 
In \cite{ICRCMinaya}, by studying the rise-time of the WCD signals at different zenith angles, there are also shown several parameters sensitive to $\langle \ln A \rangle$.  Results confirm that none of the post-LHC models produce an accurate picture of muons in air showers.

\subsection{Detector response to muons}

A dedicated Resistive Plate Chambers (RPC) hodoscope was built and installed around one of the surface detectors, namely, the  {\it Gianni Navarra} WCD, in order to study in detail its response as a function of the trajectories within the water-Cherenkov tank, and thus cross-check some of the systematic uncertainties contributing to the muon measurements \cite{ICRCAssis}. The system is able to measure single muon tracklengths on the detector water with an accuracy of   $\sim$ cm. A simulation was developed to take into account not only the geometry of the setup but also a realistic flux of atmospheric particles.

\begin{figure}[h]
  \begin{center}
    \includegraphics[width=7cm]{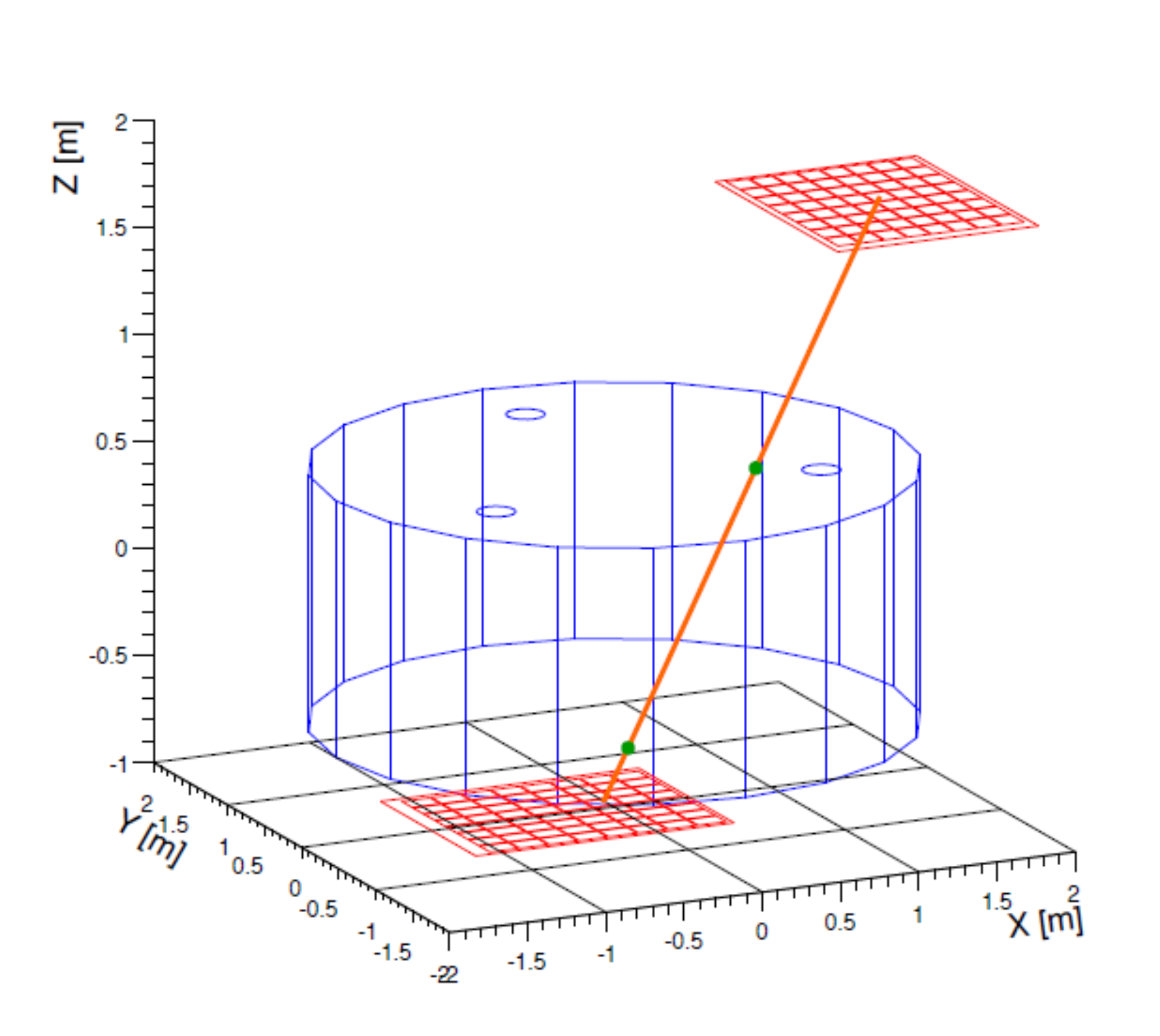}
\caption[]{Schematic view of a muon track going through the hodoscope mounted on the {\it Gianni Navarra} water-Cherenkov detector.}
\label{GN}
 \end{center}
\end{figure}

\begin{figure}[h]
  \begin{center}
    \includegraphics[width=6cm]{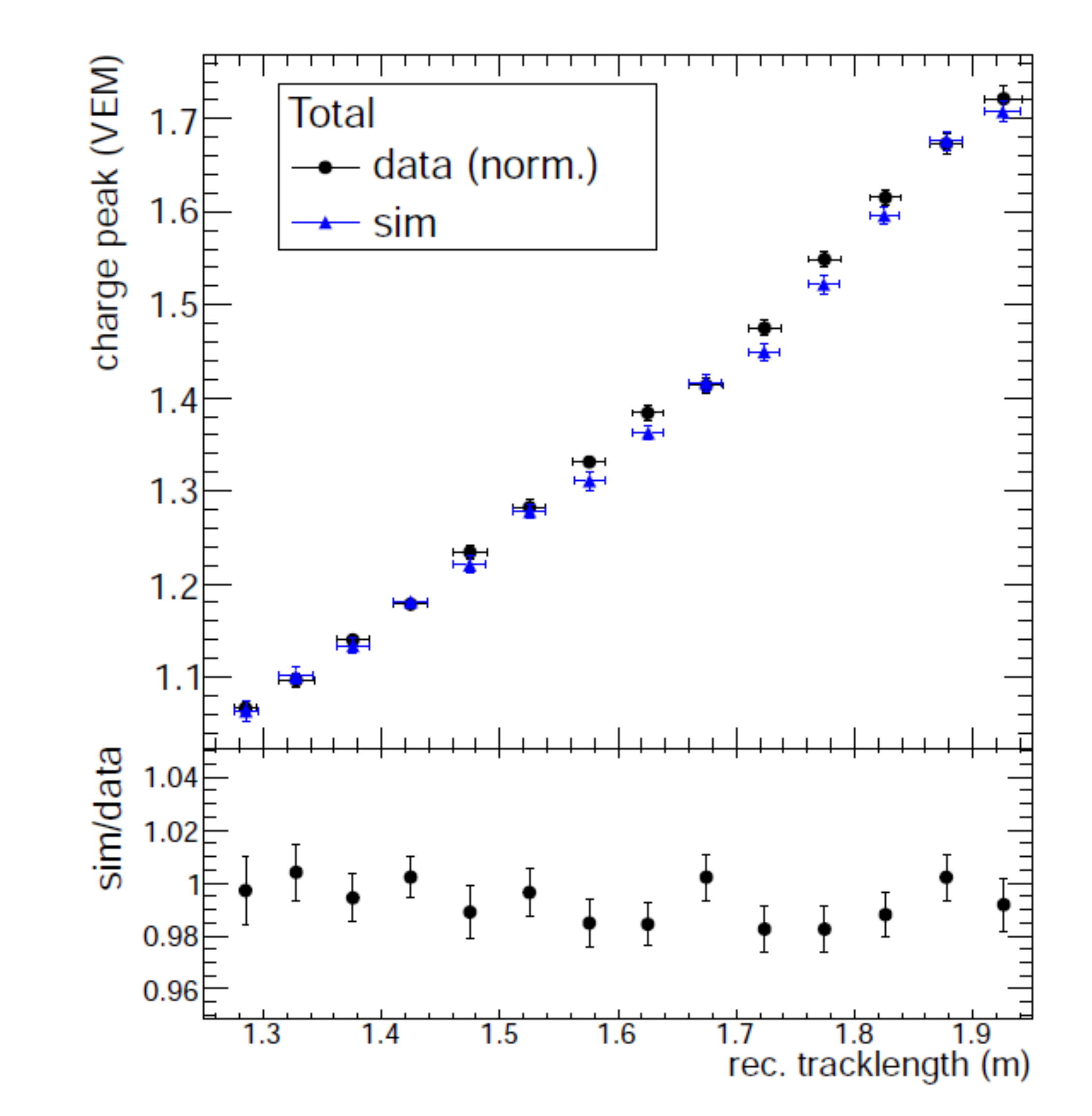}
\caption[]{Peak of the charge distribution as a function of the muon tracklength in the water-Cherenkov detector for simulations and data (normalized to simulations at the first bin). The ratio between simulations and data is also shown at the bottom.}
\label{ST}
 \end{center}
\end{figure}

Figure \ref{GN} displays a scheme of one of the setups together with a muon trajectory. Figure \ref{ST} displays the total signal as a function of the tracklength. In the current analysis,  data has been normalized to simulations at the first tracklength bin to analyze the relative evolution of trajectories from vertical to inclined ones. Data and simulations behave as expected within $\sim 2$\%.

\section{Conclusions}
\label{}
 
The Pierre Auger Observatory has updated the p-air cross section into two energy bins, both beyond the reach of LHC. A number of measurements have been also presented which have the potential to constrain hadronic interactions characteristics, by means of consistency checks of data against predictions of air shower simulations using high energy hadronic models. Related to the EM cascade,  measurements have been presented on the the average shower longitudinal profile, and also implications of the measured moments of the $\Xmax$-distributions for $\sigma^2[\ln A]$, which impose constraints in QGSJETII-04.

The analysis of the hadronic component shows several important inconsistencies on the high energy model predictions: there is a deficit in simulations on the overall number of muons, which is connected with the hadronic cascade when compared to the actual primary composition as deduced by $\Xmax$ data. The best model to describe it is EPOS-LHC, which however display an inconsistent description of the  hadronic longitudinal development (diagnosed by $\Xmumax$) for the corresponding measured $\Xmax$ values. See \cite{ICRCPierog} for a discussion about possible causes.

It is also important to note the efforts carried by different accelerator experiments (see contributions \cite{CRISUnger,CRISBongi,CRISGiacobbe} to these proceedings) to further reduce the uncertainties in the hadronic properties at low energy that could present effects on the air shower initiated at high energy. 
The Pierre Auger Observatory is  to deploy during the following years an upgrade \cite{CRISDiGiulio} aimed to improve the EM versus muonic component separation on air showers. This will allow to improve the capabilities to discriminate primary masses and also hadronic physics properties.

\section{Acknowledgments}
 

L.~Cazon wants to thank  funding by Funda\c c\~ ao para a Ci\^ encia e
Tecnologia, COMPETE, QREN,  and European Social Fund.




\nocite{*}
\bibliographystyle{elsarticle-num}
\bibliography{martin}



\end{document}